\documentclass{ws-procs975x65}

\begin{document}

\title{\uppercase{Detection of Gravitational Waves using Pulsar Timing}}

\author{\uppercase{R. N. Manchester}}

\address{CSIRO Astronomy and Space Science,\\
Australia Telescope National Facility \\
Epping, NSW 1710, Australia\\
$^*$E-mail: dick.manchester@csiro.au}

\begin{abstract}
  Pulsars are very stable clocks in space which have many
  applications to problems in physics and astrophysics. Observations
  of double-neutron-star binary systems have given the first
  observational evidence for the existence of gravitational waves
  (GWs) and shown that Einstein's general theory of relativity is an
  accurate description of gravitational interactions in the regime of
  strong gravity. Observations of a large sample of pulsars spread
  across the celestial sphere forming a ``Pulsar Timing Array'' (PTA),
  can in principle enable a positive detection of the GW background in
  the Galaxy. The Parkes Pulsar Timing Array (PPTA) is making precise
  timing measurements of 20 millisecond pulsars at three radio
  frequencies and is approaching the level of timing precision and
  data spans which are needed for GW detection. These observations
  will also allow us to establish a ``Pulsar Timescale'' and to detect
  or limit errors in the Solar System ephemerides used in pulsar
  timing analyses. Combination of PPTA data with that of other groups
  to form an International Pulsar Timing Array (IPTA) will enhance the
  sensitivity to GWs and facilitate reaching other PTA goals. The
  principal source of GWs at the nanoHertz frequencies to which PTAs
  are sensitive is believed to be super-massive binary black holes in
  the cores of distant galaxies. Current results do not signficantly
  limit models for formation of such black-hole binary systems, but in
  a few years we expect that PTAs will either detect GWs or seriously
  constrain current ideas about black-hole formation and galaxy
  mergers. Future instruments such as the Square Kilometre Array (SKA)
  should not only detect GWs from astrophysical sources but also enable
  detailed studies of the sources and the gravitational theories
  used to account for the GW emission.
\end{abstract}


\bodymatter

\section{Introduction}\label{sec:intro}
A pulsar is a rapidly rotating neutron star with very strong magnetic
fields both inside and outside the star, most likely formed in a
supernova explosion at the end of the life of a massive star. The
combination of strong magnetic field and rapid rotation generates
enormous electric fields which accelerate charged particles in the
surrounding magnetosphere to ultra-relativistic energies. These
particles radiate beams of emission which sweep across the sky as the
star rotates. If one or more of these beams cross the Earth each
rotation we can detect the neutron star as a pulsar. The interval
between pulses, known as the pulse period, is equal to the rotation
period of the star. There are currently more than 1800 pulsars
known.\footnote{See the ATNF Pulsar Catalogue \cite{mhth05} at
  www.atnf.csiro.au/research/pulsar/psrcat.}  Almost all of these lie
within our Galaxy, typically at distances of a few kiloparsec.

There are two main classes of pulsar: ``normal'' pulsars and
``millisecond'' pulsars (MSPs). Most normal pulsars have periods of
between 0.1 and 5 seconds, whereas MSPs have much shorter periods,
generally between 2 and 50 ms, hence the name. Most MSPs are binary,
that is, in orbit with another star. Orbital periods are typically
days or months, but can be as short as a few hours. Only a small
proportion of normal pulsars, just over 1\%, are binary. This gives a
pointer to the mechanism by which MSPs get their very short period --
it is believed that they were old, slow and probably dead pulsars
which were spun up by accretion of mass and angular momentum from an
evolving binary companion star (see, e.g., Ref.~\refcite{bv91}). As
well as increasing the spin frequency, the accretion evidently reduces
the effective magnetic field by several orders of magnitude. Despite
this, the increased spin rate is generally sufficient to reactivate
the emission beams -- hence the resulting short-period pulsars are
often known as ``recycled'' pulsars. Most of these are MSPs.

In general, pulsars are most easily detected at radio wavelengths and hence
most pulsar surveys are carried out in the radio band (see, e.g.,
Ref.~\refcite{mlc+01}). However, with the recent launch of the {\em
  Fermi} Gamma-ray Space Telescope, an increasing number of pulsars,
mostly young but also including many MSPs, have been found or detected
at gamma-ray wavelengths \cite{aaa+09c,aaa+09f}. At the time of
writing, the number of reported gamma-ray detected pulsars is 46.

Because neutron stars are tiny, with a radius of about 15 km, but
massive, typically about 1.4 M$_\odot$, their rotation period is
extremely stable -- they are very difficult to slow down or spin up!
It is this property which makes pulsars extraordinarily useful as
probes. This is especially true for MSPs as they have narrow pulses
(in time units) and hence measurements of pulse delays and
periodicities have higher precision compared to normal pulsars. Having
an array of precise clocks spread through the Galaxy enables many
different investigations, for example, studies of the interstellar
medium, pulsar astrometric measurements, including proper motions and
even annual parallaxes in some cases, and potentially the direct
detection of gravitational waves (GWs). The fact that most of the MSPs are
in orbit with another star is a bonus as it allows many additional
properties relating to binary motion to be investigated.

For double-neutron-star binary systems measurement of relativistic
perturbations to the orbital parameters is generally possible, thereby
allowing precise determinations of neutron star masses and making
possible stringent tests of general relativity (GR) and other
gravitational theories in the strong-field regime. The famous
Hulse-Taylor binary pulsar, PSR B1913+16, was the first binary pulsar
to be discovered \cite{ht75a}. Precise timing measurements over 30
years have given significant measurements of three relativistic or
``Post-Keplerian'' parameters, leading to accurate mass determinations
for the two stars which confirm that it is a double-neutron-star
system. They have also shown that the observed orbit decay is
consistent with the predictions of GR for energy loss from the system
in GWs, thereby providing the first observational
evidence for the existence of these waves \cite{wt05}. The Double
Pulsar, PSR J0737-3039A/B, discovered at Parkes in 2003/4
\cite{bdp+03,lbk+04}, is an even more relativistic binary system than
the Hulse-Taylor binary. Relativistic precession of the longitude of
periastron is an amazing 17$^{\rm o}$~yr$^{-1}$ and was detected in just a
few days of observation. In less than three years of observation five
Post-Keplerian parameters have been observed, which together with the
mass ratio of the two stars (measurable since both stars are detected
as pulsars), has shown that GR accurately predicts the motions of the
two stars at a level of precision exceeding 0.05\%, the most precise
test so far in this strong-field regime \cite{ksm+06}.

The main topic of this presentation is the current effort to make a
direct detection of GWs using pulsar timing. The concept of a ``Pulsar
Timing Array'' (PTA) is described in Section~\ref{sec:pta} and
Section~\ref{sec:ppta} describes the Parkes Pulsar Timing Array (PPTA)
in some detail. Section~\ref{sec:discn} discusses some of the
implications of current results and prospects for the future. Finally,
the main points are summarised in Section~\ref{sec:summary}.

\section{Pulsar Timing Arrays and Detection of Gravitational Waves}\label{sec:pta}
The existence of GWs, fluctuations in spacetime which propagate at the
speed of light, is a prediction of Einstein's general theory of
relativity and other subsequent theories of relativistic
gravitation. GWs are generated by acceleration of massive
objects. Sources of primary interest in the astrophysical context are
fluctuations in density and hence spacetime from the inflationary era
of cosmological expansion, oscillations of cosmic strings in the early
Universe, formation of massive black holes, binary super-massive black
holes in the cores of galaxies, coalescing double-neutron-star binary
systems and compact X-ray binaries in our Galaxy. Each of these
sources emits a characteristic spectrum of GWs and in principle they
can be identified by this spectrum and the time-dependence of the GW
signal. Most of the sources consist of a random accumulation of
signals from thousands or millions of individual sources and hence
form a stochastic background of GWs which we can observe.

Huge efforts have gone into making a direct detection of GWs over the
past few decades. Recently these efforts have been dominated by the
development of laser-interferometer systems such as LIGO \cite{bar00}
and VIRGO \cite{aaa+04g}. These systems have two perpendicular
interferometer arms, typically of length 3 or 4 km, and are sensitive
to GWs in the frequency range 100 -- 500 Hz. Their primary source of
interest is coalescence of double-neutron-star binary systems. They
have been operating in their ``initial'' configuration for several
years with significant upper limits on several types of source (e.g.,
Ref.~\refcite{aaa+06b,aaa+09h}). Efforts are currently under way to
develop ``advanced'' configurations which will have improved
sensitivity. For example, full operation of Advanced LIGO is expected
in 2014. Another significant project, still in the planning stage, is
LISA, a system with three spacecraft in a triangular array trailing
the Earth in its orbit about the Sun \cite{dan00}. The interferometric
arm length for LISA is about 5 million km, giving it sensitivity to
GWs in the 0.1 -- 100 mHz band. Prime targets for LISA will be the
final stages of coalescence of super-massive black-hole binary systems
in galaxy cores and Galactic X-ray binary systems. Its planned launch
date is 2020.

Gravitational waves in our Galaxy modulate the apparent period of
pulsars observed on Earth. The modulation is actually the difference
between the varying gravitational potential at the pulsar and that at
the Earth. For a stochastic background of GWs, these two modulations
are uncorrelated. With observations of just one (or a few) pulsars, a
limit can be placed on the amplitude of the GW background in the
Galaxy since any modulation due to GWs in the Galaxy cannot be much
greater than the smallest observed rms timing residual. The expected
GW spectrum from the ensemble of binary super-massive black holes in
galaxies is very ``red'', that is, stronger at low frequencies (e.g.,
Ref.~\refcite{jb03}). Also, the sensitivity of pulsar timing
experiments depends on the accumulated pulse phase offset due to the
GW signal, that is, on the integral of the pulse frequency
modulation. For both of these reasons, pulsar timing experiments are
most sensitive to long-period GW signals, with maximum sensitivity at
periods comparable to the data span, typically several years,
corresponding to GW frequencies of about 10 nHz.

Analysis of an 8-year span of Arecibo observations of PSR B1855+09 by
Kaspi, Taylor \& Ryba \cite{ktr94} placed a limit on the energy
density of the GW background relative to the closure density of the
Universe, $\Omega_{gw}$, of about $10^{-7}$ at a frequency of
approximately 1/(8 yr) or 4 nHz. More recently Jenet et
al. \cite{jhv+06} have combined the Kaspi et al. Arecibo data set with
Parkes observations of seven pulsars made as part of the PPTA project
to place the best limit so far: $\Omega_{gw}\sim 2 \times 10^{-8}$ at
$\nu_{gw} \sim$~1/(8 yr). The implications of this limit will be
discussed in Section~\ref{sec:discn}.

\begin{figure}[ht]
\begin{center}
\psfig{file=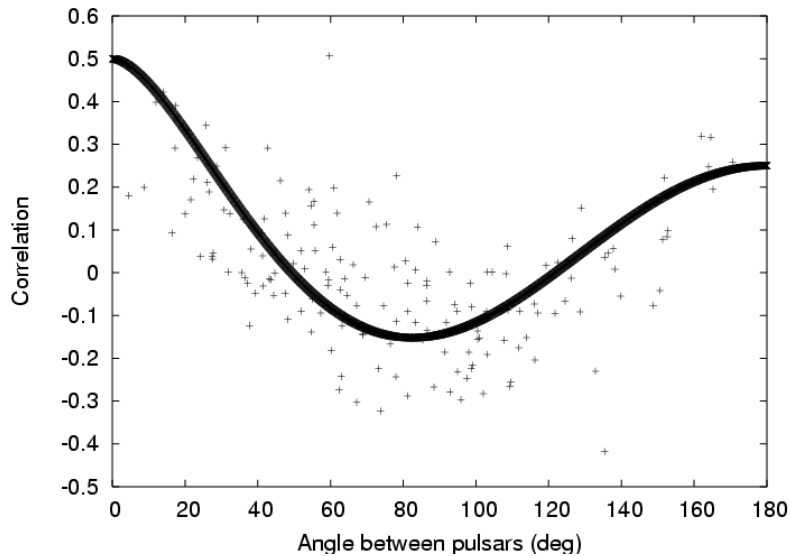,width=75mm,angle=270}
\end{center}
\caption{The expected correlation between timing residuals for pairs
  of pulsars as a function of the angle between the two pulsars for a
  stochastic GW background signal. The curve is the relation derived
  by Hellings \& Downs \cite{hd83} and the points are from a
  simulation using {\sc tempo2} with the PPTA pulsar sample where the
  GW signal dominates all other noise sources \cite{hjl+09}. }
\label{fg:hd83}
\end{figure}

Setting a limit on the strength of the GW background is of interest,
but not as interesting as making an actual detection of GWs! With
observations of many pulsars spread across the celestial sphere and
long data spans, we can in principle achieve this more ambitious
goal. Such an observational system is known as a ``Pulsar Timing Array''
(PTA). Making an actual detection of GWs relies on the fact that a GW
passing over the Earth produces a correlated signal in the residuals
of different pulsars. For a stochastic GW background signal in GR, the
correlation between the residuals for pairs of pulsars is dependent
only on the angular separation of the pulsar pair. The expected
correlation as a function of angular separation is shown in
Fig.~\ref{fg:hd83}. For pulsars which are close together on the sky
the correlation coefficient is 0.5, not 1.0 because the
modulations by GWs passing over the pulsars are uncorrelated. This
uncorrelated signal also results in the scatter in correlation for the
simulated points at similar angular separation. For pulsars which are
separated by about 90$^{\rm o}$ on the sky, the expected correlation is
negative because of the quadrupolar nature of GWs and it goes positive
again for pulsars which are more-or-less in opposite directions. 

GWs are not the only thing that can produce a correlated signal in the
timing residuals. All pulsar timing data are referenced to an
established timescale, for example, TT(TAI) or TT(BIPM09)
\cite{pet07}\footnote{See also www.bipm.org/en/scientific/tai/}. These
timescales are based on weighted averages of data from atomic
frequency standards at laboratories around the world. These frequency
standards are not perfect and hence the international timescales are
subject to various instablities. In principle, these instabilities are
detectable using pulsar timing observations, since the pulsar periods
are not subject to terrestrial influences to any significant
extent. For example, if during PTA observations all pulsars are observed
to be running slower than average, then this probably means that the
reference timescale is running fast. This signal has no spatial
signature and so can be thought of as a monopole as compared to the
quadrupolar signal from a GW. We can therefore use the pulsar data to
establish a ``Pulsar Timescale'' which, over long intervals, may be
more precise than the best timescales based on terrestrial atomic
clocks.  It is important to note that, because of the intrinsically
unknown pulsar spin frequency and slow-down rate, pulsar timing
analyses must solve for these terms thereby removing any sensitivity
to linear or quadratic terms in the GW signal and Pulsar Timescale.

We also depend on Solar System ephemerides to give the position
of the Earth relative to the Solar System barycentre at any given
time. Most pulsar analyses use ephemerides produced by NASA's Jet
Propulsion Laboratory, for example DE405 \cite{sta98b}, to derive the
barycentric corrections. These ephemerides are also not perfect and
hence the correction can be in error. In the pulsar timing analysis
this is equivalent to an error in the Earth's velocity with respect to
an inertial frame and hence shows up as a spatial dipole signature in the
timing residuals. 

Therefore in principle we can separate these three global timing
signatures by their dependence on the position of the pulsars on the
sky. In fact, both the reference timescales and the Solar System
ephemerides are both very precise and these signatures have not yet
been detected in PTA data. However, differences between successive
realisations of the timescales and the ephemerides suggest that pulsar
timing is capable of detecting errors in current versions. Similarly,
we have not yet detected GWs, but estimates of the expected GW
background amplitude and of the effect of this background on pulsar
timing data \cite{jb03,wl03a,jhlm05,svc08} suggest that with current
technology, GW signals should be detectable with 5 -- 10 years of
high-quality timing data for at least 20 MSPs. Frequent observations,
typically every 2 -- 3 weeks, are required to adequately sample the
expected signals and to improve sensitivity. Observations at two or
more radio frequencies are required to remove variable interstellar
dispersion delays.

\begin{figure}[ht]
\begin{center}
\psfig{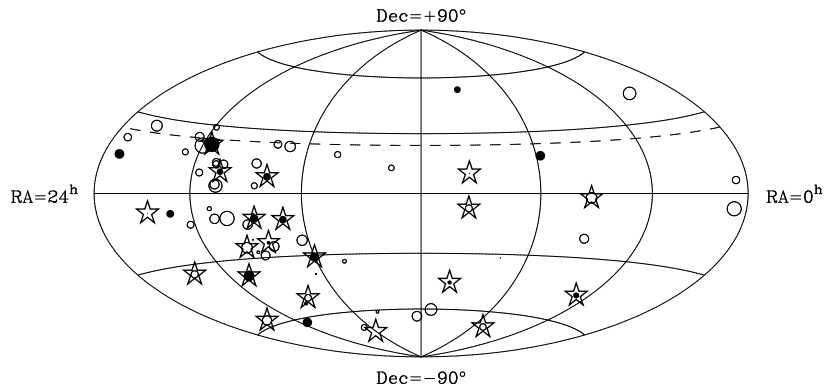}
\end{center}
\caption{The celestial distribution of pulsars suitable for PTA
  observations. The diameter of the circle is inversely related to the
  pulsar period and the circle is filled if the mean 1400 MHz flux
  density is more than 2 mJy. The dashed line is the northern
  declination limit of the Parkes radio telescope. PPTA
  pulsars are marked by a star. }
\label{fg:pta_sky}
\end{figure}

Fig.~\ref{fg:pta_sky} shows that there are about 30 MSPs known which
are strong enough and have sufficiently short periods to be suitable
for PTA projects, i.e. to give rms timing residuals of less than about
1 $\mu$s with observation times of an hour or so with existing
observing systems. Most of these are south of the celestial equator,
although hopefully on-going searches will uncover more suitable
pulsars with a wider distribution on the sky. Currently there are
three major PTA projects gathering high-quality timing data:
\begin{itemlist}
\item The European Pulsar Timing Array (EPTA)
\item The North American pulsar timing array (NANOGrav)
\item The Parkes Pulsar Timing Array (PPTA)
\end{itemlist}
The EPTA uses radio telescopes at Effelsberg, Jodrell Bank, Nan\c{c}ay
and Westerbork to obtain data. Normally observations are made
separately at the different telescopes, but there is a plan to combine
signals in real time to improve the sensitivity
\cite{fvb+10}. Currently the EPTA is obtaining good quality data for
about nine pulsars. NANOGrav uses the Arecibo and Green Bank
Telescopes and currently has good quality data for about 17 MSPs. The
PPTA uses the Parkes radio telescope and currently has good quality
data for 20 MSPs.

These three groups are collaborating to form the International Pulsar
Timing Array (IPTA) \cite{haa+10}. In the future we can expect other PTAs
to join this effort; for example, groups at the FAST and other radio
telescopes in China plan to undertake a PTA project
\cite{slk+09}. Combining data sets gives better sampling of pulsars
common to two or more PTAs and also increases the total number of
pulsars observed, giving better coverage of the celestial sphere. Both
of these things help improve our sensitivity to GWs. 

\section{The Parkes Pulsar Timing Array}\label{sec:ppta}
The PPTA project uses the Parkes 64-m radio telescope of the Australia
Telescope National Facility (ATNF) to observe 20 MSPs at 2 -- 3 week
intervals. The project is based at the ATNF and has about 25 team
members, including five students, with principal collaborating groups
at the Swinburne University of Technology (led by Matthew Bailes), the
University of Texas at Brownsville (led by Rick Jenet) and the
University of California, San Diego (led by Bill Coles). The 20 MSPs
observed by the PPTA are indicated in
Fig.~\ref{fg:pta_sky}. Observations are made at three frequencies, around
700 MHz, 1400 MHz and 3100 MHz.

Observations at 1400 MHz use the centre beam of the 20cm
Multibeam receiver \cite{swb+96} which has a bandwidth of 256 MHz and
a system equivalent flux density of about 30 Jy. Observations at the
other two bands are made simultaneously using the 10cm/50cm dual-band
receiver which has bandwidths of 64 MHz and 1024 MHz at 50cm and 10cm
respectively. System equivalent flux densities are about
64 Jy and 48 Jy, respectively, in the two bands. Data are recorded using two
different systems, polyphase digital filterbanks, which have
bandwidths up to 1024 MHz, and a baseband recording system, CPSR2,
which records two 64-MHz bands \cite{hbo06}. Data are processed using
the {\sc psrchive} programs \cite{hvm04} and the {\sc tempo2} timing
analysis program \cite{hem06}. The PPTA observing system is described
in more detail by Manchester et al. (Ref.~\refcite{mhb+10}).

Regular PPTA observations commenced in mid-2004. Since then, data
recording and analysis systems have been steadily
improved. Table~\ref{tb:ppta_psrs} lists the 20 pulsars being observed
and the current timing performance. The pulsar period and dispersion
measure (DM) are given, along with the orbital period if the pulsar is
a member of a binary system. The final column gives the rms timing
residual from a fit to approximately 1.5 years of data with one of the
more recent digital filterbank systems, PDFB2. The rms residuals refer
to summed data from a single observation, typically one hour in
duration. These results show that we are close to achieving the timing
precision that is required for a positive detection of GWs, assuming
that the predictions of the amplitude of the GW stochastic background
signal in the Galaxy are correct.

Further improvement in the precision of these arrival times can be
expected with improved signal processing and the spans of good quality
data are increasing. Analysis of 10-year spans of archival Parkes data
combined with CPSR2 data up to 2008 for the PPTA pulsars by Verbiest
et al. (Ref.~\refcite{vbc+09}) has shown that, for most of the PPTA
pulsars, there is no evidence for intrinsic period fluctuations above
the instrumental noise level over these long data spans. Although this
study needs to be extended with higher quality data over similarly
long data spans, the result is very encouraging for PTA detection of
GWs as it indicates that these efforts will not be significantly
limited by intrinsic pulsar period irregularities. Also, as mentioned
above, combining PPTA data with data from other PTAs will also improve
GW detection sensitivity. The prospects for GW detection using
pulsar timing in the next few years are good!

\begin{table}[t]
\tbl{PPTA pulsars and PDFB2 timing results}
{\begin{tabular}{lccccc}
\toprule
\multicolumn{1}{c}{PSR} & Period & DM & Orbital Period & Band & Rms Residual \\
  &  (ms)  & (cm$^{-3}$ pc) & (d) & & ($\mu$s) \\ 
\colrule
J0437$-$4715 & \hphantom{0}5.757 & \hphantom{00}2.65 & \hphantom{00}5.74 & 10cm & 0.06 \\
J0613$-$0200 & \hphantom{0}3.062 & \hphantom{0}38.78 & \hphantom{00}1.20 & 20cm & 0.54 \\
J0711$-$6830 & \hphantom{0}5.491 & \hphantom{0}18.41 & -- & 20cm & 1.27 \\
J1022+1001   & 16.453 & \hphantom{0}10.25 & \hphantom{00}7.81 & 10cm & 1.80 \\
J1024$-$0719 & \hphantom{0}5.162 & \hphantom{00}6.49 & -- & 20cm & 1.06 \\
J1045$-$4509 & \hphantom{0}7.474 & \hphantom{0}58.15 & \hphantom{00}4.08 & 20cm & 1.59 \\
J1600$-$3053 & \hphantom{0}3.598 & \hphantom{0}52.19 & \hphantom{0}14.34 & 20cm & 0.28 \\
J1603$-$7202 & 14.842 & \hphantom{0}38.05 & \hphantom{00}6.31 & 20cm & 0.96 \\
J1643$-$1224 & \hphantom{0}4.622 & \hphantom{0}62.41 & 147.02 & 20cm & 0.94 \\
J1713+0747   & \hphantom{0}4.570 & \hphantom{0}15.99 & \hphantom{0}67.83 & 10cm & 0.20 \\
J1730$-$2304 & \hphantom{0}8.123 & \hphantom{00}9.61 & -- & 20cm & 1.62 \\
J1732$-$5049 & \hphantom{0}5.313 & \hphantom{0}56.84 & \hphantom{00}5.26 & 20cm & 2.89 \\
J1744$-$1134 & \hphantom{0}4.075 & \hphantom{00}3.14 & -- & 10cm & 0.41 \\
J1824$-$2452 & \hphantom{0}3.054 & 119.86 & -- & 20cm & 1.95 \\
J1857+0943   & \hphantom{0}5.362 & \hphantom{0}13.31 & \hphantom{0}12.33 & 20cm & 0.45 \\
J1909$-$3744 & \hphantom{0}2.947 & \hphantom{0}10.39 & \hphantom{00}1.53 & 10cm & 0.11 \\
J1939+2134   & \hphantom{0}1.558 & \hphantom{0}71.04 & -- & 10cm & 0.17 \\
J2124$-$3358 & \hphantom{0}4.931 & \hphantom{00}4.62 & -- & 20cm & 2.86 \\
J2129$-$5721 & \hphantom{0}3.726 & \hphantom{0}31.85 & \hphantom{00}6.63 & 20cm & 1.49 \\
J2145$-$0750 & 16.052 & \hphantom{00}9.00 & \hphantom{00}6.84 & 20cm & 0.36 \\
\botrule
\end{tabular}}
\label{tb:ppta_psrs}
\end{table}

\section{Discussion}\label{sec:discn}
\begin{figure}[ht]
\begin{center}
\psfig{file=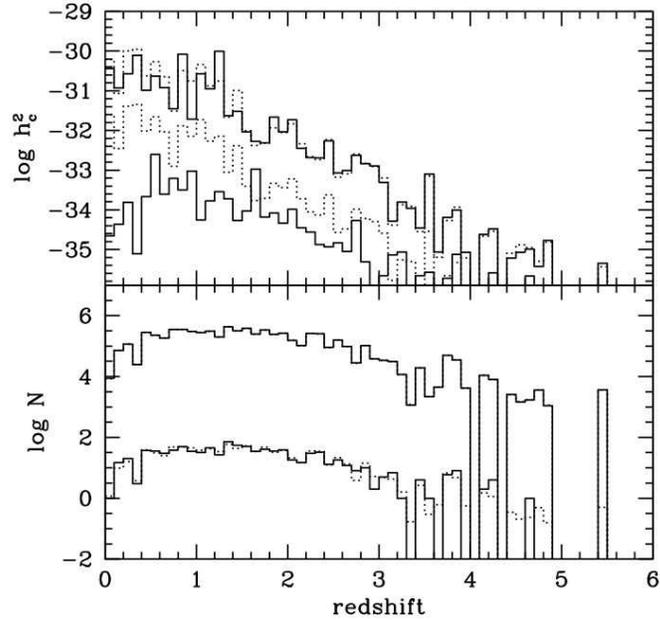,width=90mm}
\end{center}
\caption{Contribution of different redshift intervals to the
    stochastic GW background for two GW frequencies, $8\times
    10^{-9}$~Hz (upper curve in each subplot) and $10^{-7}$~Hz (lower
    curve in each subplot). The upper panel shows the signal amplitude
    and the lower panel shows the number of galaxies contributing. The
    solid lines are from a Monte Carlo method and the dashed lines are
    from a semi-analytic approach. (From Ref.~\refcite{svc08}.)}
\label{fg:gw_sim}
\end{figure}

Current PTAs are most sensitive to GWs in the frequency range 3 -- 10
nHz. The astrophysical GW signal in this frequency range most likely
to be detected is a stochastic background from super-massive
black-hole binary systems in the cores of distant galaxies
\cite{jb03,wl03a}. Simulations of the expected GW signal by Sesana,
Vecchio \& Colacino (Ref.~\refcite{svc08}) show that it is dominated
by radiation from galaxies with intermediate redshifts $z\sim 1$
(Fig.~\ref{fg:gw_sim}) and that the strongest signals come from
systems with black-hole masses of $10^9$ -- $10^{10}$ M$_\odot$.

\begin{figure}[ht]
\begin{center}
\psfig{file=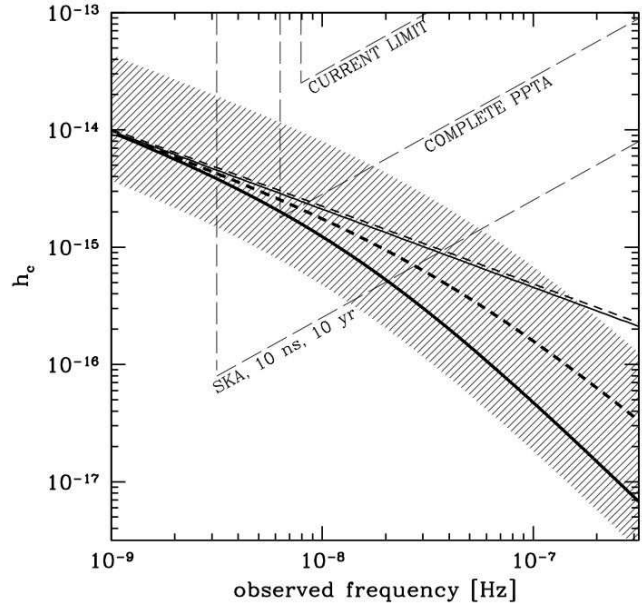,width=90mm}
\end{center}
\caption{The range of predictions of the strain amplitude as a
  function of GW frequency from different models is represented by the
  hashed area. The lines within this region are predictions from
  particular models. The long-dashed lines are the approximate
  sensitivity limits for PTAs, with the upper one being the current
  best published limit, the middle one being an ``ideal'' PPTA with 20
  pulsars, 100 ns rms timing residuals and a 5-year data span, and the
  lower one is a prediction of the limit obtainable with the
  proposed Square Kilometer Array radio telescope after 10
  years. (From Ref.~\refcite{svc08}.) }
\label{fg:gw_amp}
\end{figure}

Fig.~\ref{fg:gw_amp} shows the range of model predictions for the
strain amplitude as a function of GW frequency. Current published
limits (e.g., Ref.~\refcite{jhv+06}) do not constrain the model, but
the ``ideal'' PPTA or, more realistically, longer data spans with
current instrumentation and combined IPTA data sets will either detect
the stochastic GW background or significantly modify our current
understanding of black-hole formation in galaxy cores and/or galaxy
merger processes. Future instruments such as the Square Kilometre
Array (SKA) will have enormously improved sensitivity and are expected
to obtain high-precision TOAs for hundreds of MSPs \cite{kbc+04}. With
a data span of five years or more, this will certainly give a
significant detection of GWs unless the current predictions are totally
wrong. It will enable not just a detection of GWs, but also detailed
study of the signal. For example, if one of several non-Einsteinian
theories of gravitation is correct, then GWs will have polarization
properties which differ from those predicted by GR and this will have
an effect on the observed correlations \cite{ljp08,cjp09}. Even
current results are constraining some proposed sources of the GW
background. For example, the results presented by Jenet et
al. (Ref.~\refcite{jhv+06}) limit the equation of state of matter at
the epoch of inflation, $w = P/\epsilon > -1.3$
(cf. Ref.~\refcite{gri05}) and the tension in cosmic strings
(cf. Ref.~\refcite{dv05}).

\begin{figure}[ht]
\begin{center}
\psfig{file=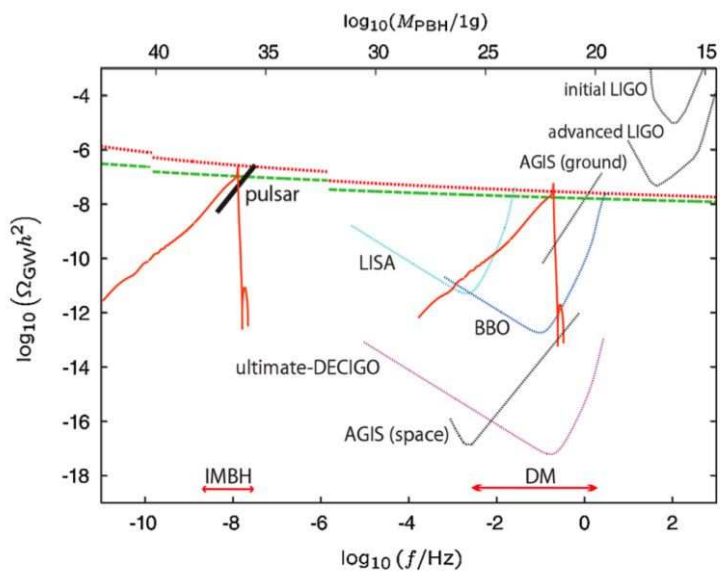,width=100mm}
\end{center}
\caption{Spectrum of GWs emitted by formation of black holes at the
  end of the inflation era (Ref.~\refcite{sy09}). For a given
  formation event, the emitted spectrum is concentrated at a GW
  frequency which is inversely related to the collapsing mass as
  indicated on the upper x-scale. Spectra are shown for two specific
  black-hole masses: 600 M$_\odot$ (IMBH, peak around $10^{-8}$~Hz in
  the PTA frequency range) and $1.5\times 10^{-11}$~M$_\odot$ (Dark
  Matter, peak around 0.1~Hz in the LISA frequency range). These
  spectra peak near the dotted and dashed lines which are envelopes
  for the GW emssion from this mechanism with two different sets of
  assumptions.  Sensitivity limits for other proposed or existing
  instruments are also shown.  Current pulsar limits, indicated by the
  thick black line, already severely limit IMBH formation by this
  process. (See Ref.~\refcite{sy09} for more details.)}
\label{fg:imbh}
\end{figure}

Intermediate-mass black holes (IMBHs), with masses in the range of a few 100
to a few 1000 M$_\odot$, have been invoked as the basis of the
so-called ultra-luminous X-ray sources (see Ref.~\refcite{mc04a} for a
review). IMBHs are difficult to form by aggregation of
stellar-mass objects. It has been suggested by Saito \& Yokoyama
(Ref.~\refcite{sy09}) that the most likely formation path is by
collapse of overdense regions at the end of the inflation era. Saito
\& Yokoyama show that formation of such IMBHs at this time will lead to
a potentially detectable background of GWs. Fig.~\ref{fg:imbh} shows
the spectrum of the emitted GWs. The GW frequency is dependent on the
mass of the formed black hole and is in the PTA range for
IMBHs. In fact, the predicted amplitude of the GW background from IMBH
collapse is greater than the limits already placed by pulsar timing,
making primordial formation of IMBHs unlikely. This raises the issues
of whether or not such IMBHs exist, what the energy source of
ultra-luminous X-ray sources is and maybe even whether or not
these X-ray sources are actually ultra-luminous.

\begin{figure}[ht]
\begin{center}
\psfig{file=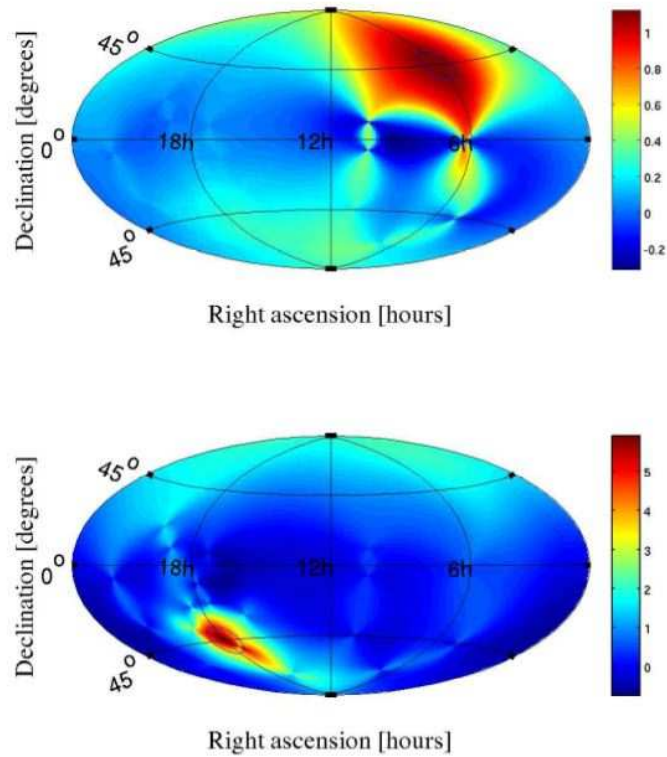,width=90mm}
\end{center}
\caption{Localisation of a source of GWs in the northern sky (upper) and in
the southern sky (lower) using simulated data for the pulsars observed
by the PPTA. The actual assumed source positions are 06$^{\rm h}$, 45$^{\rm
  o}$ and 18$^{\rm h}$, $-45^{\rm o}$. (Ref.~\refcite{abc+09})}
\label{fg:sky_posn}
\end{figure}

It is possible that future PTAs will have sufficient sensitivity to
detect individual sources of GWs.  Sesana, Vecchio \& Volonteri
(Ref.~\refcite{svv09}) show that current models for the evolution of
super-massive binary black holes in galaxies predict that at least one
such system will produce a sinusoidal signal of period comparable to
the data span with an amplitude of between 5 and 50 ns in timing
residuals. A signal of this size is potentially detectable with the
IPTA and certainly should be detectable by the SKA. Such a detection
would open up the field of GW astronomy, allowing investigation of the
source characteristics and the properties of the GW emission
itself. For example, it will be possible to localise the source on the
sky. Fig.~\ref{fg:sky_posn} shows a simulation of the ability of the
PPTA to determine source positions \cite{abc+09}. Since the PPTA
pulsars are predominantly in the southern hemisphere, source positions
are better determined in the south with a point-spread function of
half-power width about 15$^{\rm o}$ or 20$^{\rm o}$. An improved sky
distribution of pulsars and higher sensitivity from the IPTA or SKA
will improve the accuracy to which a source can be localised on the
sky.

\section{Summary}\label{sec:summary}
Pulsars are extremely stable clocks distributed throughout our
Galaxy. The high precision to which it is possible to measure pulse
arrival times at the Earth and the long data spans now available leads
to many interesting applications. For example, pulsar timing has given
the first observational evidence for the existence of gravitational
waves and shown that Einstein's general theory of relativity is
accurate in the regime of strong gravitational fields. With
observations of many pulsars distributed across the celestial sphere
-- a ``Pulsar Timing Array'' -- we can in principle make a positive
detection of gravitational waves from astrophysical sources. Such an
array can also be used to establish a pulsar-based timescale and to
identify (or limit) errors in Solar-system ephemerides. 

\begin{figure}[ht]
\begin{center}
\psfig{file=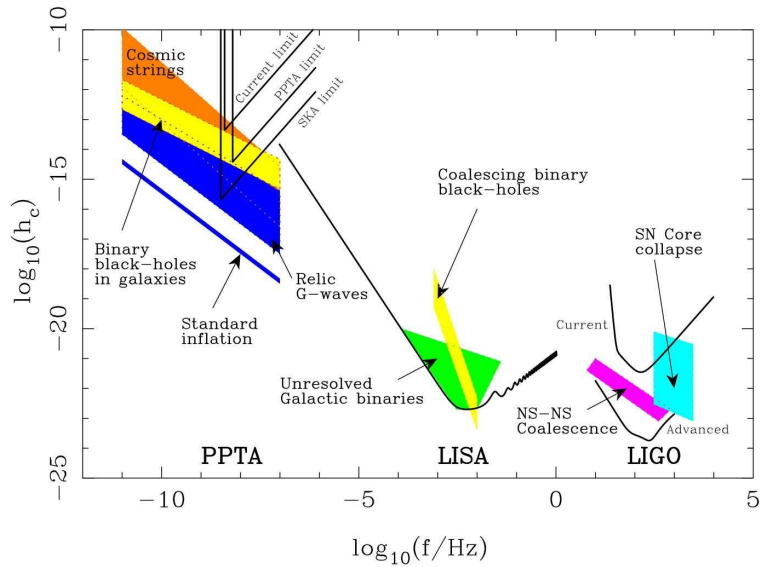,width=120mm}
\end{center}
\caption{Spectrum of potentially detectable GW sources and sensitivity
curves for PTA systems, the space-based laser interferometer LISA and
the ground-based laser interferometer LIGO. }
\label{fg:gw_spec}
\end{figure}

We are now approaching the level of timing precision and data spans
which are needed to achieve the main goals of PTA
projects. Specifically, if current predictions of the strength of the
stochastic GW background in the Galaxy from binary super-massive black
holes in the cores of distant galaxies are realistic, within a few years
we should be able to make a significant detection of this
signal. Other possible sources of a stochastic GW background include
fluctuations in the inflation era and oscillations of cosmic strings
in the early Universe. Fig.~\ref{fg:gw_spec} summarises the signals
expected from different astrophysical sources and places the PTA in
the context of ground- and space-based laser interferometer systems,
specifically LIGO and LISA. 

Current efforts are concentrating on eliminating systematic
errors from the timing data and improving signal-processing and
GW-detection algorithms. While intrinsic period irregularities are
significant in a few PTA pulsars, so far they are not a limiting
factor for GW detection. Several groups around the world are
collaborating to combine data sets to form an International Pulsar
Timing Array, which will improve our sensitivity to GWs and help us to
reach the other
PTA goals. Future instruments such as the Square Kilometre Array will
greatly enhance the sensitivity of these efforts and will surely allow
the detection and detailed study of GWs and their sources, both
stochastic and individual. These studies have the
potential to give new information on the properties of gravitational
waves and gravitational interactions in the near and distant Universe. 

\section{Acknowledgments}
I thank all members of the PPTA collaboration and the staff of Parkes
Observatory; without their efforts this presentation
would not be possible. The Parkes radio telescope is part of the Australia
Telescope which is funded by the Commonwealth Government for operation
as a National Facility managed by CSIRO.


\end{document}